\documentclass[aps,prl,preprint,superscriptaddress]{revtex4-1}

\usepackage[english]{babel}
\usepackage[utf8]{inputenc} 
\usepackage{graphicx}
\usepackage{bm}
\usepackage{siunitx}
\usepackage[colorlinks=true, pdfstartview=FitV, linkcolor=blue,
citecolor=blue, urlcolor=blue]{hyperref}

\begin{document}


\title{Imaging stray magnetic field of individual ferromagnetic
  nanotubes}


\author{D. Vasyukov} \thanks{Authors contributed equally.}
\affiliation{Department of Physics, University of Basel, 4056 Basel,
  Switzerland}
		
\author{L. Ceccarelli} \thanks{Authors contributed equally.}
\affiliation{Department of Physics, University of Basel, 4056 Basel,
  Switzerland}
		
\author{M. Wyss} \affiliation{Department of Physics, University of
  Basel, 4056 Basel, Switzerland}

\author{B. Gross} \affiliation{Department of Physics, University of
  Basel, 4056 Basel, Switzerland}
  
\author{A. Schwarb} \affiliation{Department of Physics, University of
  Basel, 4056 Basel, Switzerland}
  
\author{A. Mehlin} \affiliation{Department of Physics, University of
  Basel, 4056 Basel, Switzerland}

\author{N. Rossi} \affiliation{Department of Physics, University of
  Basel, 4056 Basel, Switzerland}
	
\author{G. T\"{u}t\"{u}nc\"{u}oglu} \affiliation{Laboratory of
  Semiconductor Materials, Institute of Materials (IMX), School of
  Engineering, \'Ecole Polytechnique F\'ed\'erale de Lausanne (EPFL),
  1015 Lausanne, Switzerland}
  
\author{F. Heimbach} \affiliation{Lehrstuhl f\"{u}r Physik
  funktionaler Schichtsysteme, Physik Department E10, Technische
  Universit\"{a}t M\"{u}nchen, 85747 Garching, Germany}
	
\author{R. R. Zamani} \affiliation{Solid State Physics, Lund
  University, 22100 Lund, Sweden}
	
\author{A. Kov\'acs} \affiliation{Ernst Ruska-Centre for Microscopy
  and Spectroscopy with Electrons, Forschungszentrum Jülich, 52425
  Jülich, Germany}
	
\author{A. Fontcuberta i Morral} \affiliation{Laboratory of
  Semiconductor Materials, Institute of Materials (IMX), School of
  Engineering, \'Ecole Polytechnique F\'ed\'erale de Lausanne (EPFL),
  1015 Lausanne, Switzerland}
	
\author{D. Grundler} \affiliation{Laboratory of Nanoscale Magnetic
  Materials and Magnonics, Institute of Materials (IMX), School of
  Engineering, \'Ecole Polytechnique F\'ed\'erale de Lausanne (EPFL),
  1015 Lausanne, Switzerland}
	
\author{M. Poggio} \affiliation{Department of Physics, University of
  Basel, 4056 Basel, Switzerland} \email{martino.poggio@unibas.ch}
\homepage{http://poggiolab.unibas.ch/}



\date{\today}

\begin{abstract} 
  We use a scanning nanometer-scale superconducting quantum
  interference device to map the stray magnetic field produced by
  individual ferromagnetic nanotubes (FNTs) as a function of applied
  magnetic field.  The images are taken as each FNT is led through
  magnetic reversal and are compared with micromagnetic simulations,
  which correspond to specific magnetization configurations.  In
  magnetic fields applied perpendicular to the FNT long axis, their
  magnetization appears to reverse through vortex states, i.e.\
  configurations with vortex end domains or -- in the case of a
  sufficiently short FNT -- with a single global vortex.  Geometrical
  imperfections in the samples and the resulting distortion of
  idealized mangetization configurations influence the measured
  stray-field patterns.
\end{abstract} 


\maketitle

As the density of magnetic storage technology continues to grow,
engineering magnetic elements with both well-defined remnant states
and reproducible reversal processes becomes increasingly challenging.
Nanometer-scale magnets have intrinsically large surface-to-volume
ratios, making their magnetization configurations especially
susceptible to roughness and exterior imperfections.  Furthermore,
poor control of surface and edge domains can lead to complicated
switching processes that are slow and not reproducible
reproducible~\cite{zheng_switching_1997, fruchart_enhanced_1999}.

One approach to address these challenges is to use nanomagnets that
support remnant flux-closure configurations.  The resulting absence of
magnetic charge at the surface reduces its role in determining the
magnetic state and can yield stable remnant configurations with both
fast and reproducible reversal processes.  In addition, the lack of
stray field produced by flux-closure configurations suppresses
interactions between nearby nanomagnets.  Although the stability of
such configurations requires dimensions significantly larger than the
dipolar exchange length, the absence of dipolar interactions favors
closely packed elements and thus high-density
arrays~\cite{han_nanoring_2008}.

On the nanometer-scale, core-free geometries such as
rings~\cite{lopez-diaz_computational_2000, rothman_observation_2001}
and tubes~\cite{landeros_equilibrium_2009} have been proposed as hosts
of vortex-like flux-closure configurations with magnetization pointing
along their circumference.  Such configurations owe their stability to
the minimization of magnetostatic energy at the expense of exchange
energy.  Crucially, the lack of a magnetic core removes the dominant
contribution to the exchange energy, which otherwise compromises the
stability of vortex states.

Here, we image the stray magnetic field produced by individual
ferromagnetic nanotubes (FNTs) as a function of applied field using a
scanning nanometer-scale superconducting quantum interference device
(SQUID).  These images show the extent to which flux closure is
achieved in FNTs of different lengths as they are driven through
magnetic reversal.  By comparing the measured stray-field patterns to
the results of micromagnetic simulations, we then deduce the
progression of magnetization configurations involved in magnetization
reversal.

Mapping the magnetic stray field of individual FNTs is challenging,
due to their small size and correspondingly small magnetic moment.
Despite a large number of theoretical studies discussing the
configurations supported in FNTs~\cite{hertel_magnetic_2004,
  escrig_phase_2007, escrig_effect_2007, landeros_reversal_2007,
  landeros_equilibrium_2009, chen_magnetization_2010,
  landeros_domain_2010, chen_magnetization_2011, yan_chiral_2012},
experimental images of such states have so far been limited in both
scope and detail.  Cantilever
magnetometry~\cite{weber_cantilever_2012, gross_dynamic_2016}, SQUID
magnetometry~\cite{buchter_reversal_2013, buchter_magnetization_2015},
and magnetotransport measurements~\cite{ruffer_magnetic_2012,
  baumgaertl_magnetization_2016} have recently shed light on the
magnetization reversal process in FNTs, but none of these techniques
yield spatial information about the stray field or the configuration
of magnetic moments.  Li et al.\ interpreted the nearly vanishing
contrast in a magnetic force microscopy (MFM) image of a single FNT in
remnance as an indication of a stable global vortex state, i.e.\ a
configuration dominated by a single azimuthally-aligned
vortex~\cite{li_template-based_2008}.  Magnetization configurations in
rolled-up ferromagnetic membranes between 2 and 16 $\mu$m in diameter
have been imaged using magneto-optical Kerr
effect~\cite{streubel_magnetic_2014}, x-ray transmission
microscopy~\cite{streubel_magnetic_2014}, x-ray magnetic dichroism
photoemission electron microscopy
(XMCD-PEEM)~\cite{streubel_imaging_2014}, and magnetic soft x-ray
tomography~\cite{streubel_retrieving_2015}.  More recently, XMCD-PEEM
was used to image magnetization configurations in FNTs of different
lengths~\cite{wyss_imaging_2017,stano_imaging_2017}.  Due to technical
limitations imposed by the technique, measurement as a function of
applied magnetic field was not possible.

\begin{figure}[t]
  \includegraphics{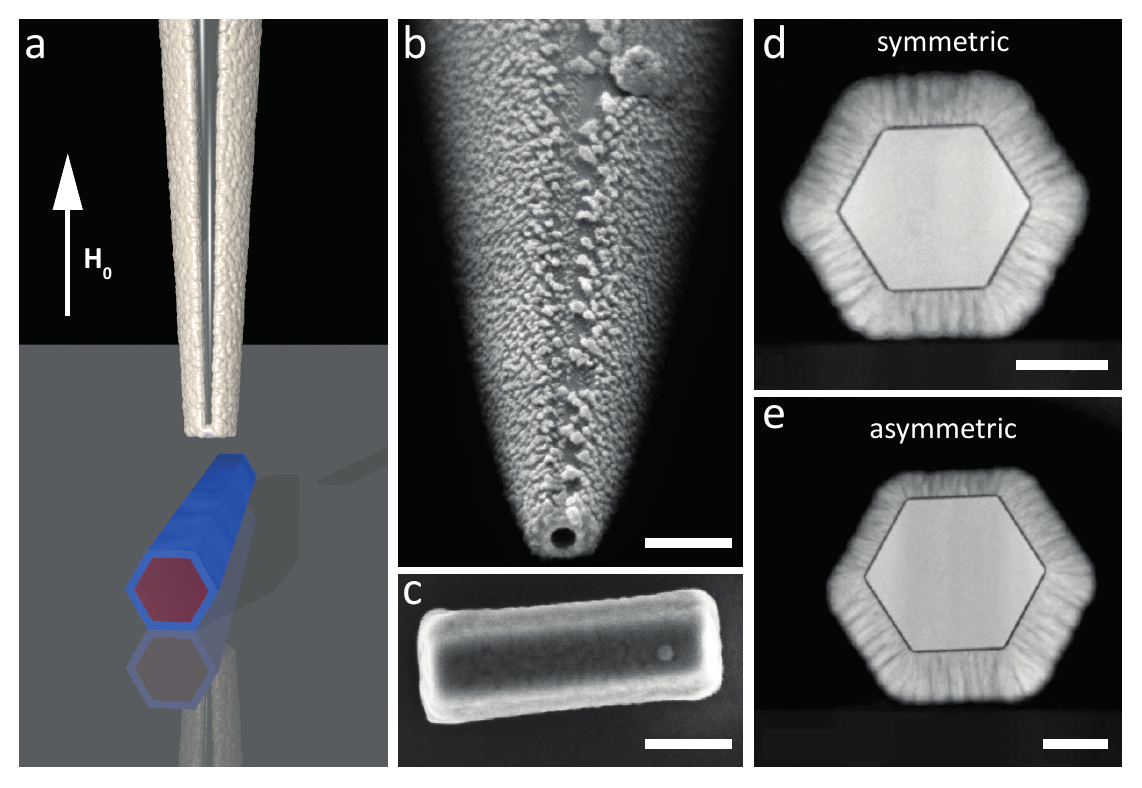}%
  \caption{Experimental setup.  (a) Schematic drawing showing the
    scanning SOT, a FNT lying on the substrate, and the direction of
    $\mathbf{H_0}$.  The CoFeB shell is depicted in blue and the GaAs
    core in red.  Pb on the SOT is shown in white.  SEMs of the (b)
    the SOT tip and (c) a 0.7-$\mu$m-long FNT.  (d) and (e) show
    cross-sectional HAADF STEMs of two FNTs from a similar growth
    batch as those measured.  The scalebars represent 200 nm in (b)
    and (c) and 50 nm in (d) and (e).}%
\label{fig1}
\end{figure}
  
We use a scanning SQUID-on-tip (SOT) sensor to map the stray field
produced by FNTs as a function of position and applied field.  We
fabricate the SOT by evaporating Pb on the apex of a pulled quartz
capillary according to a self-aligned method pioneered by Finkler et
al.\ and perfected by Vasyukov et al.~\cite{finkler_self-aligned_2010,
  vasyukov_scanning_2013}.  The SOT used here has an effective
diameter of 150 nm, as extracted from measurements of the critical
current $I_{\text{SOT}}$ as a function of a uniform magnetic field
$\mathbf{H_0} = H_0 \hat{z}$ applied perpendicular to the SQUID loop.
At the operating temperature of 4.2 K, pronounced oscillations of
critical current are visible as a function of $H_0$ up to 1 T.  The
SOT is mounted in a custom-built scanning probe microscope operating
under vacuum in a $^4$He cryostat.  Maps of the magnetic stray field
produced by individual FNTs are made by scanning the FNTs lying on the
substrate in the $xy$-plane 300 nm below the SOT sensor, as shown
schematically in Fig.~\ref{fig1}~(a).  The current response of the
sensor is proportional to the magnetic flux threaded through the SQUID
loop.  For each value of the externally applied field $H_0$, a factor
is extracted from the current-field interference pattern to convert
the measured current $I_{SOT}$ to the flux.  The measured flux then
represents the integral of the $z$-component of the total magnetic
field over the area of the SQUID loop.  By subtracting the
contribution of $H_0$, we isolate the $z$-component of stray field,
$H_{dz}$ integrated over the area of the SOT at each spatial position.

FNT samples consist of a non-magnetic GaAs core surrounded by a
30-nm-thick magnetic shell of CoFeB with hexagonal cross-section.
CoFeB is magnetron-sputtered onto template GaAs nanowires (NWs) to
produce an amorphous and homogeneous shell~\cite{gross_dynamic_2016},
which is designed to avoid magneto-crystalline
anisotropy~\cite{hindmarch_interface_2008,ruffer_anisotropic_2014,
  schwarze_magnonic_2013}. Nevertheless, recent magneto-transport
experiments show that a small growth-induced magnetic anisotropy may
be present~\cite{baumgaertl_magnetization_2016}.  Scanning electron
micrographs (SEMs) of the studied FNTs, as in Fig.~\ref{fig1}~(c),
reveal continuous and defect-free surfaces, whose roughness is less
than 5 nm.  Figs.~\ref{fig1}~(d) and (e) show cross-sectional
high-angle annular dark-field (HAADF) scanning transmission electron
micrographs (STEM) of two FNTs from the same growth batch as those
measured, highlighting the possibility for asymmetry due to the
deposition process.  Dynamic cantilever magnetometry measurements of
representative FNTs show $\mu_0 M_S = 1.3 \pm 0.1$
T~\cite{gross_dynamic_2016}, where $\mu_0$ is the permeability of free
space and $M_S$ is the saturation magnetization.  Their diameter $d$,
which we define as the diameter of the circle circumscribing the
hexagonal cross-section, is between 200 and 300 nm.  Lengths from 0.7
to 4 $\mu$m are obtained by cutting individual FNTs into segments
using a focused ion beam (FIB).  After cutting, the FNTs are aligned
horizontally on a patterned Si substrate.  All stray-field
progressions are measured as functions of $H_0$, which is applied
perpendicular to the substrate and therefore perpendicular to the long
axes of each FNT.  Gross et al.\ found that similar CoFeB FNTs are
fully saturated by a perpendicular field for $|\mu_0 H_0| > 1.2$~T at
$T = 4.2$~K~\cite{gross_dynamic_2016}.  Since the superconducting
SQUID amplifier used in our measurement only allows measurements for
$|\mu_0 H_0| \leq 0.6$~T, all the progressions measured here represent
minor hysteresis loops.

\begin{figure}[t]
  \includegraphics[width=137.25mm]{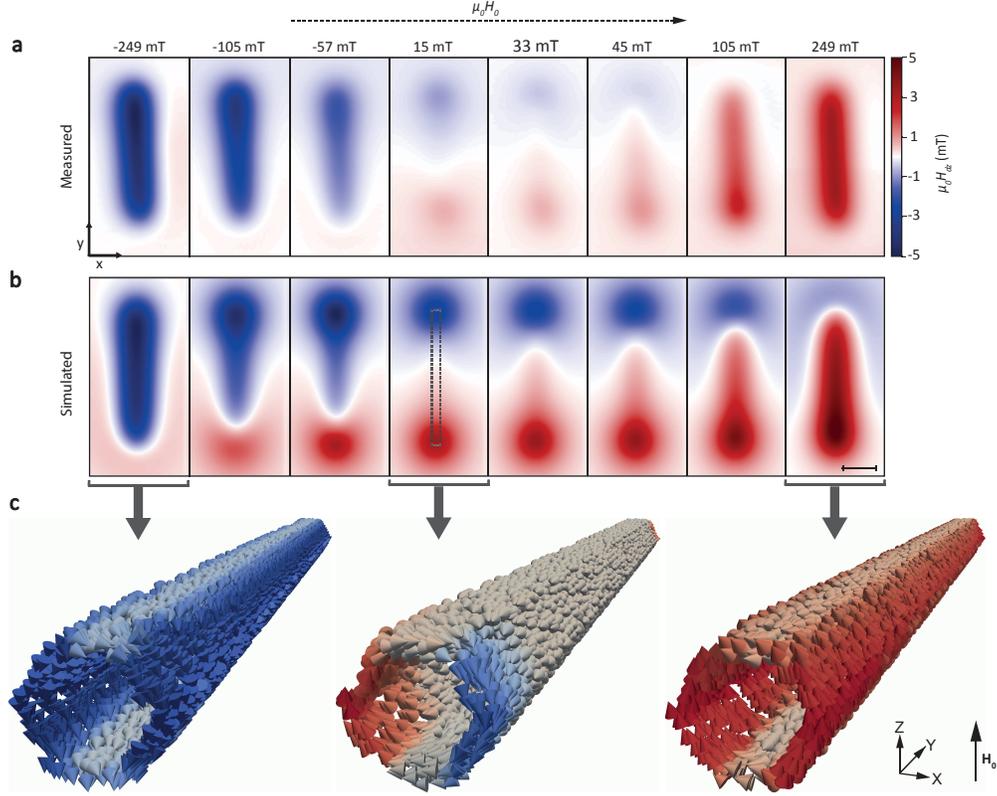}%
  \caption{Magnetic reversal of a 4-$\mu$m-long FNT ($l =
    4.08$~$\mu$m, $d = 260$~nm) in a field $H_0$ applied perpendicular
    to its long axis.  Images of the stray field component along
    $\hat{z}$, $H_{dz}$, in the $xy$-plane 300 nm above the FNT for
    the labeled values of $\mu_0 H_0$ (a) as measured by the scanning
    SOT and (b) as generated by numerical simulations of the
    equilibrium magnetization configuration.  The dashed line
    deliniates the position of the FNT.  The scalebar corresponds to
    1~$\mu$m.  (c) Simulated configurations corresponding to three
    values of $H_0$.  The middle configuration, nearest to zero field,
    shows a mixed state with vortex end domains of opposing
    circulation sense.  Arrows indicate the direction of the
    magnetization, while red (blue) contrast corresponds to the
    magnetization component along $\hat{z}$ ($-\hat{z}$).}%
\label{fig2}
\end{figure}
  
Fig.~\ref{fig2}~(a) shows the stray field maps of a 4-$\mu$m-long FNT
for a series of fields as $\mu_0 H_0$ is increased from -0.6 to 0.6~T.
The maps reveal a reversal process roughly consistent with a rotation
of the net FNT magnetization.  At $\mu_0 H_0 = -249$~mT and at more
negative fields , $H_{dz}$ is nearly uniform above the FNT, indicating
that its magnetization is initially aligned along the applied field
and thus parallel to $-\hat{z}$.  As the field is increased toward
positive values, maps of $H_{dz}$ show an average magnetization
$\langle \mathbf{M} \rangle$, which rotates toward the long axis of
the FNT.  Near $H_0 = 0$, the two opposing stray field lobes at the
ends of the FNT are consistent with an $\langle \mathbf{M} \rangle$
aligned along the long axis.  With increasing positive $H_0$, the
reversal proceeds until the magnetization aligns along $\hat{z}$.

The simulated stray-field maps, shown in Fig.~\ref{fig2}~(b), are
generated by a numerical micromagnetic model of the equilibrium
magnetization configurations.  We use the software package
\textit{Mumax3}~\cite{vansteenkiste_design_2014}, which employs the
Landau-Lifshitz micromagnetic formalism with finite-difference
discretization.  The length $l = 4.08$~$\mu$m and diameter $d =
260$~nm of the FNT are determined by SEMs of the sample, while the
thickness $t = 30$~nm is taken from cross-sectional TEMs of samples
from the same batch.  As shown in Fig.~\ref{fig2}, the simulated
stray-field distributions closely match the measurements.  The
magnetization configurations extracted from the simulations are
non-uniform, as shown in Fig.~\ref{fig2}~(c).  In the central part of
the FNT, the magnetization of the different facets in the hexagonal
FNT rotates separately as a function of $H_0$, due to their shape
anisotropy and their different orientations.  As $H_0$ approaches
zero, vortices nucleate at the FNT ends, resulting in a low-field
mixed state, i.e.\ a configuration in which magnetization in the
central part of the FNT aligns along its long axis and curls into
azimuthally-aligned vortex domains at the ends.  Experimental evidence
for such end vortices has recently been observed by
XMCD-PEEM~\cite{wyss_imaging_2017} and
DCM~\cite{mehlin_observation_2017} measurements of similar FNTs at
room-temperature.  We also measured and simulated a 2-$\mu$m-long FNT
of similar cross-sectional dimensions.  It shows an analogous
progression of stray field maps as a function of $H_0$ (see
supplementary material).  Simulations suggest a similar progression of
magnetization configurations, with a mixed state in remnance.

\begin{figure}[t]
  \includegraphics[width=139.7mm]{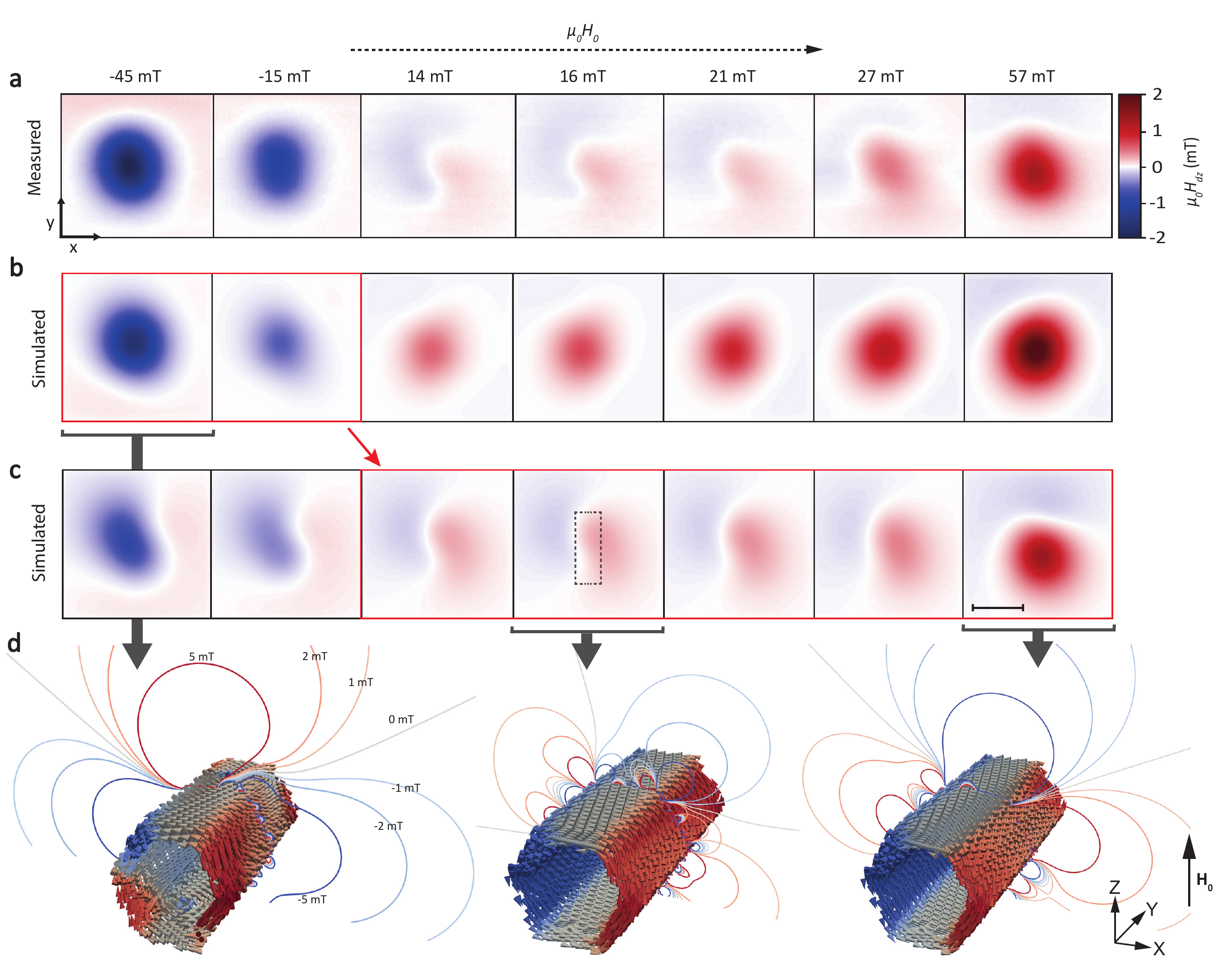}%
  \caption{Magnetic reversal of a 0.7-$\mu$m-long FNT ($l =
    0.69$~$\mu$m, $d = 250$~nm) in a field applied perpendicular to
    its long axis.  Images of the stray field component along
    $\hat{z}$, $H_{dz}$, in the $xy$-plane 300 nm above the FNT for
    the labeled values of $H_0$ (a) as measured by the scanning SOT.
    (b) and (c) show numerical simulations of $H_{dz}$ produced by two
    progressions of equilibrium magnetization configurations with
    different initial conditions.  The dashed line deliniates the
    position of the FNT and the scalebar corresponds to 0.5~$\mu$m.
    (c) Magnetization configurations and contours of constant $H_{dz}$
    corresponding to three values of $H_0$. The configuration on the
    left is characterized by two vortices in the top and bottom
    facets, respectively.  The middle and left configurations are
    distorted global vortex states.  Arrows indicate the direction of
    the magnetization, while red (blue) contrast corresponds to the
    magnetization component along $\hat{z}$ ($-\hat{z}$).}%
\label{fig3}
\end{figure}
  
FNTs shorter than 2~$\mu$m exhibit qualitatively different stray-field
progressions.  Measurements of a 0.7-$\mu$m-long FNT are shown in
Fig.~\ref{fig3}~(a).  A stray-field pattern with a single lobe
persists from large negative field to $\mu_0 H_0 = -15~\text{mT}$
without an indication of $\langle \mathbf{M} \rangle$ rotating towards
the long axis.  Near zero field, a stray-field map characterized by an
'S'-like zero-field line appears (white contrast in
Fig.~\ref{fig3}~(a)).  At more positive fields, a single lobe again
dominates.  A similar progression of stray field images is also
observed upon the reversal of a 1-$\mu$m-long FNT (not shown).

In order to infer the magnetic configuration of the FNT, we simulate
its equilibrium configuration as a function of $H_0$ using the
sample's measured parameters: $l = 0.69$~$\mu$m, $d = 250$~nm and $t =
30$~nm.  For a perfectly hexagonal FNT with flat ends, the simulated
reversal proceeds through different, slightly distorted global vortex
states, which depend on the initial conditions of the magnetization.
Such simulations do not reproduce the 'S'-like zero-field line
observed in the measured stray-field maps.  However, when we consider
defects and structural asymmetries likely to be present in the
measured FNT, the simulated and measured images come into agreement.

In these refined simulations, we first consider the magnetic
'dead-layer' induced by the FIB cutting of the FNT ends as previously
reported~\cite{katine_patterning_2003,khizroev_focused-ion-beam-based_2004,knutson_magnetic_2008}.
We therefore reduce the length of the simulated FNT by 100 nm on
either side.  Second, we take into account that the FIB-cut ends of
the FNT are not perfectly perpendicular to its long axis.  SEMs of the
investigated FNT show that the FIB cutting process results in ends
slanted by $10^\circ$ with respect to $\hat{z}$.  Finally, we consider
that the 30-nm-thick hexagonal magnetic shell may be asymmetric, i.e.\
slightly thicker on one side of the FNT due to an inhomogeneous
deposition, e.g.\ Fig.~\ref{fig1}~(e).

With these modifications, the simulated reversal proceeds through at
least four different possible stray-field progressions depending on
the initial conditions.  Only two of these, shown in
Figs.~\ref{fig3}~(b) and (c), produce stray-field maps which resemble
the measurement.  The measured stray-field images are consistent with
the series shown in Fig.~\ref{fig3}~(b) for negative fields ($\mu_0
H_0 = -45, -15~\text{mT}$).  As the applied field crosses zero ($
-15~\text{mT} \leq \mu_0 H_0 \leq 14~\text{mT}$), the FNT appears to
change stray-field progressions.  The images taken at positive fields
($ 14~\text{mT} \leq \mu_0 H_0$), show patterns consistent with the
series shown in Fig.~\ref{fig3}~(c).  The magnetic configurations
corresponding to these simulated stray-field maps suggest that the FNT
occupies a slightly distorted global vortex state.  Before entering
this state, e.g.\ at $\mu_0 H_0 = -45~\text{mT}$, the simulations show
a more complex configuration with magnetic vortices in the top and
bottom facets, rather than at the FNT ends.  On the other hand, at
similar reverse fields, e.g.\ $\mu_0 H_0 = 57~\text{mT}$, the FNT is
shown to occupy a distortion of the global vortex state with an tilt
of the magnetization toward the FNT long axis in some of the hexagonal
facets.

\begin{figure}[t]
  \includegraphics[width=81mm]{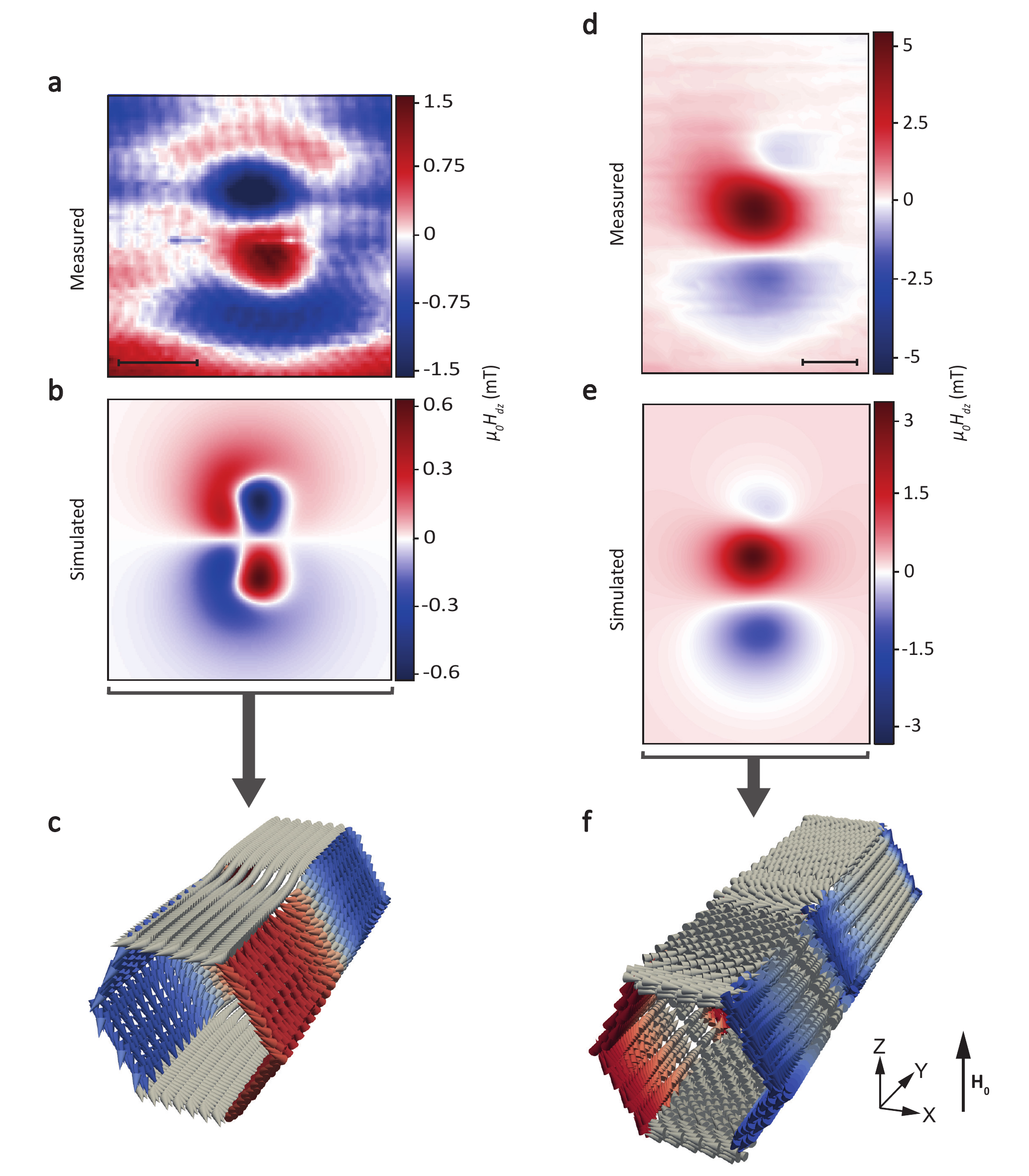}%
  \caption{Anomalous stray-field patterns found at low applied field.
    (a) Stray-field pattern of the 0.7-$\mu$m-long FNT ($l =
    0.69$~$\mu$m, $d = 250$~nm) at $\mu_0 H_0 = 20~\text{mT}$.  (b)
    Similar map produced by an opposing vortex state, shown
    schematically in (c) and observed near zero field by Wyss et
    al.~\cite{wyss_imaging_2017}.  (d) Stray-field pattern of the
    1-$\mu$m-long FNT ($l = 1.05$~$\mu$m, $d = 250$~nm) at $\mu_0 H_0
    = 21~\text{mT}$.  (e) Similar field map produced by a (f)
    multi-domain mixed state with vortex end domains and opposing
    axial domains separated by a vortex wall.  The scalebar
    corresponds to 0.5~$\mu$m.  In (c) and (f), arrows indicate the
    direction of the magnetization, while red (blue) contrast
    corresponds to the magnetization component along $\hat{z}$}%
\label{fig4}
\end{figure}

For some minor loop measurements of short FNTs ($l \leq 1$ $\mu$m), we
obtain stray-field patterns, which the micromagnetic simulations do
not reproduce.  Two such cases are shown in Fig.~\ref{fig4}, where (a)
represents the stray-field pattern measured above a 0.7-$\mu$m-long
FNT at $\mu_0 H_0 = 20~\text{mT}$ and (d) the pattern measured above a
1-$\mu$m FNT at $\mu_0 H_0 = 21~\text{mT}$.  Both of these stray-field
maps are qualitatively different from the results of Fig.~\ref{fig3}.
Since the simulations do not provide equilibrium magnetization
configurations that generate these measured stray-field patterns, we
test a few idealized configurations in search of possible matches.  In
particular, the measured pattern shown in Fig.~\ref{fig4}~(a) is
similar to the pattern produced by an opposing vortex state.  This
configuration, shown in Fig.~\ref{fig4}~(c), consists of two vortices
of opposing circulation sense, separated by a domain wall.  It was
observed with XMCD-PEEM to occur in similar-sized
FNTs~\cite{wyss_imaging_2017} in remnance at room-temperature.  The
pattern measured in Fig.~\ref{fig4}~(e) appears to match the
stray-field produced by a multi-domain state consisting of two
head-to-head axial domains separated by a vortex domain wall and
capped by two vortex ends, shown in Fig.~\ref{fig4}~(f).  Although
these configurations are not calculated to be equilibrium states for
these FNTs in a perpendicular field, they have been suggested as
possible intermediate states during reversal of axial magnetization in
a longitudinal field~\cite{landeros_reversal_2007}.  The presence of
these anomalous configurations in our experiments may be due to
incomplete magnetization saturation or imperfections not taken into
account by our numerical model.

Wyss et al.\ showed that the types of remnant states that emerge in
CoFeB FNTs depend on their length~\cite{wyss_imaging_2017}.  For FNTs
of these cross-sectional dimensions longer than 2~$\mu$m, the
equilibrium remnant state at room temperature is the mixed state,
while shorter FNTs favor global or opposing vortex states.  Here, we
confirm these observations at cyrogenic temperatures by mapping the
magnetic stray-field produced by the FNTs rather than their
magnetization.  In this way, we directly image the defining property
of flux-closure configurations, i.e.\ the extent to which their stray
field vanishes.  In fact, we find that the imperfect geometry of the
FNTs causes even the global vortex state to produce stray fields on
the order of 100~$\mu$T at a distance of 300~nm.  Finer control of the
sample geometry is required in order to reduce this stray field and
for such devices be considered as elements in ultra-high density
magnetic storage.  Using the scanning SQUID's ability to make images
as a function of applied magnetic field, we also reveal the
progression of stray-field patterns produced by the FNTs as they
reverse their magnetization.  Future scanning SOT experiments in
parallel applied fields could further test the applicability of
established theory to real
FNTs~\cite{usov_domain_2007,landeros_reversal_2007,landeros_equilibrium_2009,landeros_domain_2010}.
While the incomplete flux closure and the presence of magnetization
configurations not predicted by simulation indicate that FNT samples
still cannot be considered ideal, scanning SOT images show the promise
of using geometry to program both the overall equilibrium
magnetization configurations and the reversal process in nanomagnets.

\textbf{Methods.}  \textit{SOT Fabrication.}  SOTs were fabricated
according to the technique described by Vasyukov et
al.~\cite{vasyukov_scanning_2013} using a three-step evaporation of Pb
on the apex of a quartz capillary, pulled to achieve the required SOT
diameter.  The evaporation was performed in a custom-made evaporator
with a base pressure of $2 \times 10^{-8}$ mbar and a rotateable
sample holder cooled by liquid He.  In accordance with Halbertal et
al.~\cite{halbertal_nanoscale_2016}, an additional Au shunt was
deposited close to the tip apex prior to the Pb evaporation for
protection of the SOTs against electrostatic discharge.  SOTs were
characterized in a test setup prior to their use in the scanning probe
microscope.

\textit{SOT Positioning and Scanning.}  Positioning and scanning of
the sample below the SOT is carried out using piezo-electric
positioners and scanners (Attocube AG).  We use the sensitivity of the
SOT to both temperature and magnetic
field~\cite{halbertal_nanoscale_2016} in combination with electric
current, which is passed through a serpentine conductor on the
substrate, to position specific FNTs under the SOT (see supplementary
material).

\textit{FNT Sample Preparation.}  The template NWs, onto which the
CoFeB shell is sputtered, are grown by molecular beam epitaxy on a Si
(111) substrate using Ga droplets as catalysts
\cite{ruffer_anisotropic_2014}.  During CoFeB sputter deposition, the
wafers of upright and well-separated GaAs NWs are mounted with a
35$^\circ$ angle between the long axis of the NWs and the deposition
direction.  The wafers are then continuously rotated in order to
achieve a conformal coating.  In order to obtain NTs with different
lengths and well-defined ends, we cut individual NTs into segments
using a Ga FIB in a scanning electron microscope.  After cutting, we
use an optical microscope equipped with precision micromanipulators to
pick up the FNT segments and align them horizontally onto a Si
substrate.  FNT cross-sections for the HAADF STEMs were also prepared
using a FIB.

\textit{Mumax3 Simulations.}  To simulate the CoFeB FNTs, we set
$\mu_0 M_S$ to its measured value of $\SI{1.3}{\tesla}$ and the
exchange stiffness to $A_{ex} = \SI{28}{\pico\joule}/\si{\meter}$.
The external field is intentionally tilted by $2^\circ$ with respect
to $\hat{z}$ in both the $xz$- and the $yz$-plane, in order to exclude
numerical artifacts due to symmetry.  This angle is within our
experimental alignment error.  The asymmetry in the magnetic
cross-section of an FNT, seen in Fig.~\ref{fig1}~(e), is generated by
removing a hexagonal core from a larger hexagonal wire, whose axis is
slightly shifted.  In this case, the wire's diameter is 30~nm larger
than the core's diameter and we shift the core's axis below that of
the wire by 5~nm.  In order to rule out spurious effects due to the
discretization of the numerical cells, the cell size must be smaller
than the ferromagnetic exchange length of 6.5~nm.  This criterion is
fulfilled by using a 5-nm cell size to simulate the 0.7-$\mu$m-long
FNT.  For the 4-$\mu$m-long FNT, computational limitations force us to
set the cell size to 8~nm, such that the full scanning field can be
calculated in a reasonable amount of time.  Given that the cell size
exceeds the exchange length, the results are vulnerable to numerical
artifacts.  To confirm the reliability of these simulations, we
perform a reference simulation with a 4-nm cell size.  Although the
magnetic states are essentially unchanged by the difference in cell
size, the value of the stray field is altered by up to 10\%.


\begin{acknowledgments}
  \noindent We thank Jordi Arbiol and Rafal Dunin-Borkowski for work
  related to TEM, Sascha Martin and the machine shop of the Department
  of Physics at the University of Basel for technical support, and
  I. Dorris for helpful discussions.  We acknowledge the support of
  the Canton Aargau, ERC Starting Grant NWScan (Grant No. 334767), the
  SNF under Grant No. 200020-159893, the Swiss Nanoscience Institute,
  the NCCR Quantum Science and Technology (QSIT), and the DFG
  Schwerpunkt Programm ``Spincaloric transport phenomena'' SPP1538 via
  Project No. GR1640/5-2.
\end{acknowledgments}

\bibliography{sSOTCoFeB}

\end{document}